\documentclass[twocolumn,twoside,slac_two]{revtex4}
\usepackage{graphicx}
\usepackage{fancyhdr}
\usepackage{subfigure}
\usepackage{amsmath}
\usepackage{xspace}

\newcommand{\mett}{\mbox{${E\!\!\!\!/_T}$}}

\newcommand{\NONE}{\mbox{$\tilde{\chi}_1^0$}}

\newcommand{\HT}{H_{T}}

\newcommand{\DPHI}{\mbox{$\Delta\phi(\gamma_{1}, \gamma_{2})$}}
\def\Gravitino{\tilde{G}}
\def\bi{\begin{itemize}}
\def\ei{\end{itemize}}
\def\bc{\begin{center}}
\def\ec{\end{center}}
\def\and{\/\mbox{and}}

\newcommand{\grav}{\ensuremath{\tilde{G}}}

\newcommand{\none}{\NONE}

\begin{document}

\title{Exotic Photon Searches at CDF II}

%

\author{Eunsin Lee (for the CDF Collaboration)}
\affiliation{Department of Physics, Texas A\&M University, College Station, TX
77843, USA}

\begin{abstract}
We present recent results of searches for exotic photons at CDF~II. In the first
signature-based search, we search for anomalous production of two photons with additional
energetic objects. The results are consistent with the standard model
expectations. In the second analysis, we present a signature-based search for
anomalous production of events containing a photon, two jets, of which at least
one is identified as originating from a $b$ quark, and missing transverse
energy. We find no indications of non-standard model phenomena. Finally, a
search for a fermiophobic Higgs in the diphoton final state is presented. Since
no evidence of a resonance in the diphoton mass spectrum is observed we exclude
this Higgs boson with mass below 106~GeV/$c^2$ at a 95\% confidence level.
\end{abstract}

\maketitle

\thispagestyle{fancy}


\section{Introduction}
Over the last decades, the fast developments in phenomenology and model-building
have high-energy physicists at the Tevatron with a number of new physics
scenarios to investigate. Searches at CDF have been either broad signature-based
searches in accessible data samples for any discrepancy with the standard model
(SM) in event yields or a specific new physics model-based searches. The
signature-based searches proceed quickly in unprejudiced way as well as cover
many new physics models. The model-based searches are highly sensitive for a
particular model, and provide limits on the model. In this report we present two
results using the signature-based search approach and one result in a
model-based approach. All of these analyses use photons in the final state.
\section{Search for Anomalous Production of $\gamma\gamma+X$}
We define a ``baseline'' sample with two isolated, central ($0.05<|\eta|<1.05$)
photons with $E_{T}>13$~GeV. We then select subsamples which also contain at
least one more energetic, isolated and well-defined object or where two photons
are accompanied by large missing transverse energy (\mett). The additional
object may be an electron ($e$), muon ($\mu$), $\tau$-lepton ($\tau$), or jets.
The integrated luminosity for each subsample varies from 1 to 2 fb$^{-1}$.
In next subsections we address each $\gamma\gamma+X$ ($X=e/\mu, \tau$ and \mett)
subsample in turn. We describe the definition of the subsamples, the calculation
of the SM predictions, and the comparison of the data and the predictions.
Unless it is otherwise noted, all analyses use the same definitions of the
additional objects and kinematic variables: electrons, muons, $\tau$-leptons,
jets, soft unclustered energy, \mett, and $\HT$. The $\HT$ is defined as a
scalar sum of \mett\ and $E_{T}$'s of all identified photons, leptons, and jets.

\subsection{The $\gamma\gamma+e/\mu$ Final State}
\begin{figure*}[htbp]
  \centering
  \subfigure[]{
    \includegraphics[width=.48\linewidth]{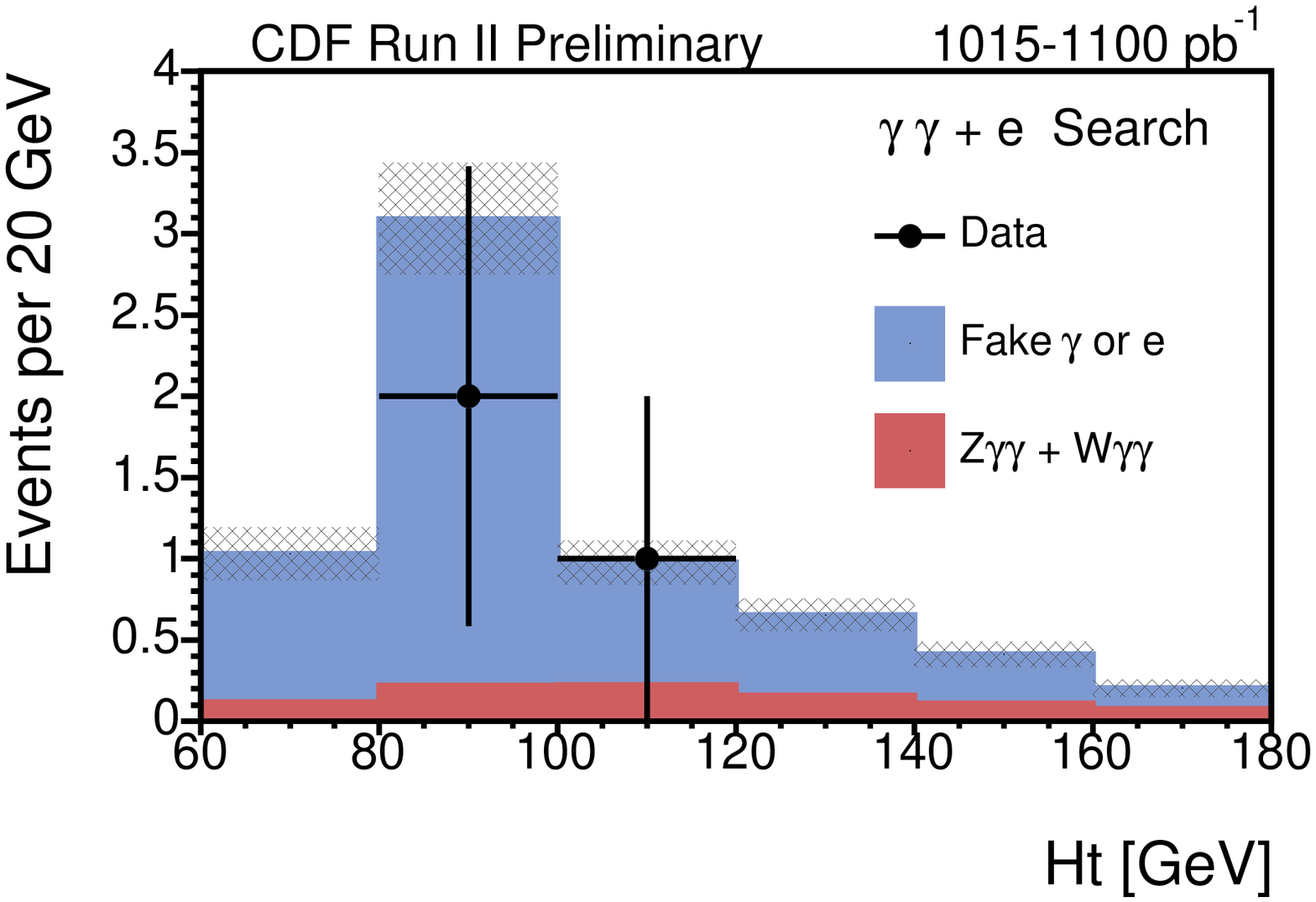}}
  \subfigure[]{
    \includegraphics[width=.48\linewidth]{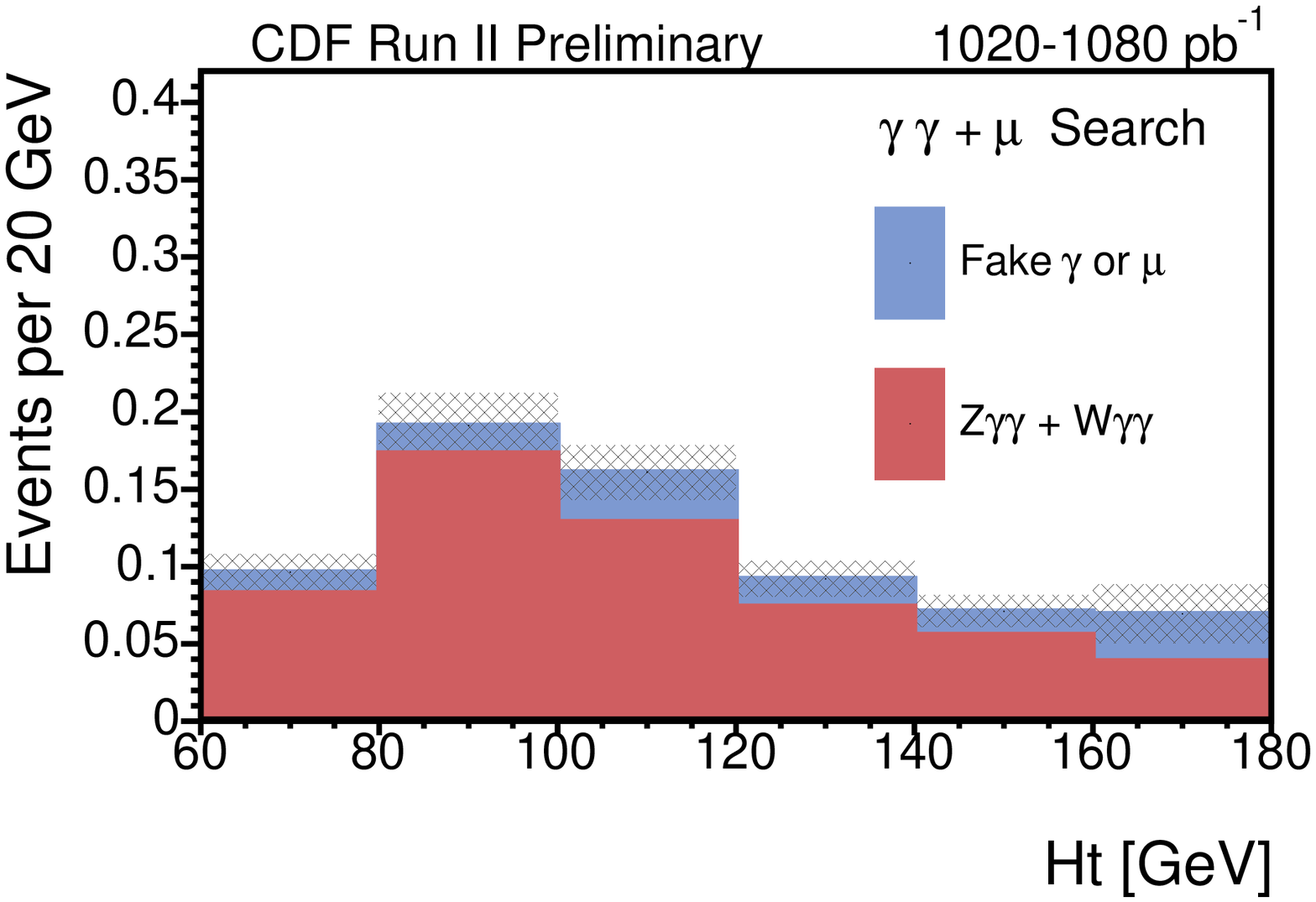}}
  \caption{In (a) the $\HT$ distributions of the $\gamma\gamma e$ events from the SM
 prediction and the three events observed in the data. In (b) the $\HT$ distributions of the $\gamma\gamma\mu$ events with zero observed events.}
  \label{fig:ggemu}
\end{figure*}

We search in 1.1 fb$^{-1}$ of data for anomalous production of events containing
two photons and at least one additional electron or muon. The selected
$\gamma\gamma e$ and $\gamma\gamma\mu$ events must have at least one electron
(central or forward) or
muon ($|\eta|<1.0$) candidate with $E^{e}_{T}>20$~GeV and $p^{\mu}_{T}>20$~GeV/$c$,
respectively. 
Backgrounds for the $\gamma\gamma e$ and $\gamma\gamma\mu$ signatures of new
physics include the SM production of $Z\to l^{+}l^{-}$ and $W\to l\bar{\nu}_{l}$
in association with two photons ($Z\gamma\gamma$, $W\gamma\gamma$), where
photons are radiated from either the initial state quarks, charged electroweak
boson ($W$), or the final-state leptons. Also there are misidentification
backgrounds (fake photons or leptons).
Backgrounds for the $\gamma\gamma e$ 
channel is dominated by $Z\gamma$ production with an electron being
misidentified as a photon. This is estimated by defining a sample of events with
two electrons and one photon, then applying a probability, which is derived in
data, for an electron to be misidentified as a photon. We find three
$\gamma\gamma e$ events with
an expected SM backgrounds of 6.82$\pm$0.75. In the $\gamma\gamma\mu$ channel
the leading background is the electroweak tri-boson production of
$Z\gamma\gamma$. We find no $\gamma\gamma\mu$ events and expect 0.79$\pm$0.11
events. Figure~\ref{fig:ggemu} shows the $\HT$ distributions from data and the
predicted backgrounds and we do not see any evidence for anomalous production of
$\gamma\gamma e$ and $\gamma\gamma\mu$ events.

\subsection{The $\gamma\gamma+\tau$ Final State}
\begin{figure}[htbp]
  \centering
 {
\includegraphics[width=1.\linewidth]{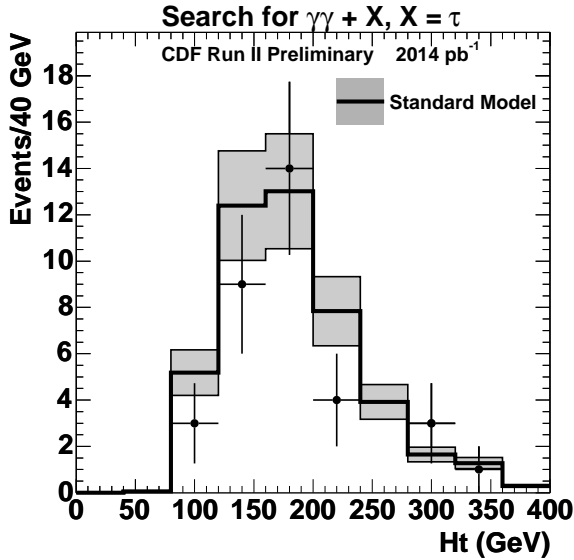}}
  \caption{The $\HT$ distribution in $\gamma\gamma+\tau$ candidate events and the
SM backgrounds.}
  \label{fig:ggtau}
\end{figure}

We search for 2.0 fb$^{-1}$ of data for events with two photons and a
hadronically decaying $\tau$-lepton. The selected
$\gamma\gamma\tau$ events must have at least one $\tau$-lepton candidate
identified using the tight requirements and passing $E_{T}>15$~GeV. We consider
two sources of backgrounds: the SM production of $W\to\tau\nu$ or $Z\to\tau\tau$
with photons and $\gamma\gamma$ events with jets misidentified as
$\tau$-leptons.
The dominant background in this search is from $\gamma\gamma+jets$ events where one of the
jets is misidentified as a $\tau$-lepton. To estimate this background, we select
events with two photons and a ``loose'' $\tau$-lepton candidate and apply the
$jet\to\tau$ misidentification probability. Since the misidentification
probability is different for jets originated by quarks or by gluons, and the
ratio of quark jets to gluon jets may be different than in the sample used to
derive the $jet\to\tau$ misidentification probability, we corrected for this
effect. We observe 34 events with 46$\pm$10
expectation. Figure~\ref{fig:ggtau} also shows the $\HT$ distribution for the
selected $\gamma\gamma+\tau$ candidate events and the predicted SM background,
which indicates there is no anomalies.

\subsection{The $\gamma\gamma+\mett$ Final State}
\begin{figure}[htbp]
  \centering
 {
\includegraphics[width=1.\linewidth]{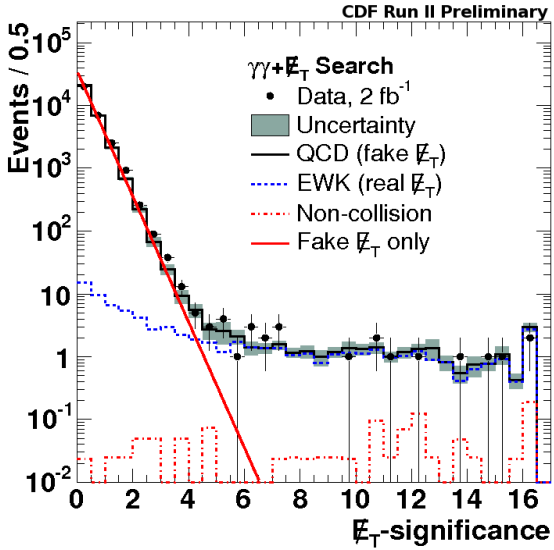}}
  \caption{The \mett-significance distribution in all $\gamma\gamma$ candidate
events from the baseline sample The QCD backgrounds with fake \mett\ are well
separated from the EWK backgrounds with real \mett.}
  \label{fig:metsig}
\end{figure}

We search for the anomalous production of two photons and large missing
transverse energy (\mett) in 2 fb$^{-1}$ of data. The \mett\ is defined as an
energy imbalance in the calorimeter and is an experimental signature of neutrino
or new weakly interacting particle. The \mett, however, can be mimicked by a
simple energy misreconstruction in SM events (``fake'' \mett): for example, fluctuations in jet
energy measurements. A better separation between events with real and fake
\mett\ can be achieved if a significance of the measured \mett\ is considered
rather than its absolute value. The \mett-significance is a dimensionless
quantity based on the energy resolution of jets and soft unclustered energy,
taking into account the event topology.
As shown in Fig.~\ref{fig:metsig}, the \mett-significance distributions have
very different shapes in events with fake and real \mett: exponentially falling
(solid line) and almost flat shapes, respectively. Thus, the \mett-significance
is an efficient tool in separating such events.
For example, a cut on the \mett-significance$>5$ which reduces the
mismeasured-energy background by a factor of $10^{5}$, the sample becomes
dominated by $W\gamma$ production, where the electron is misidentified as a
photon. This background is estimated from Monte Carlo (MC) normalized to data.
We observed 23 events and and expectation of 27.3$\pm$2.3 events.

\begin{figure}[tbp]
  \centering
\subfigure[]{
     \includegraphics[width=1.\linewidth]{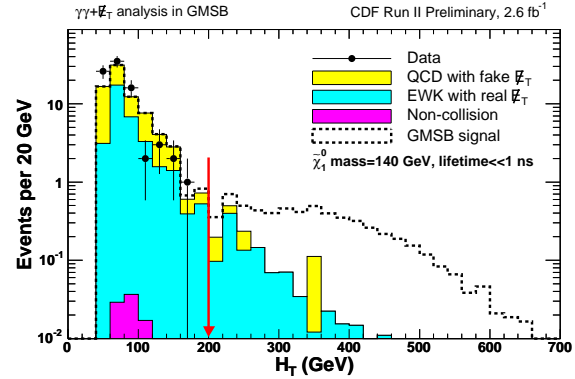}}
\subfigure[]{
    \includegraphics[width=1.\linewidth]{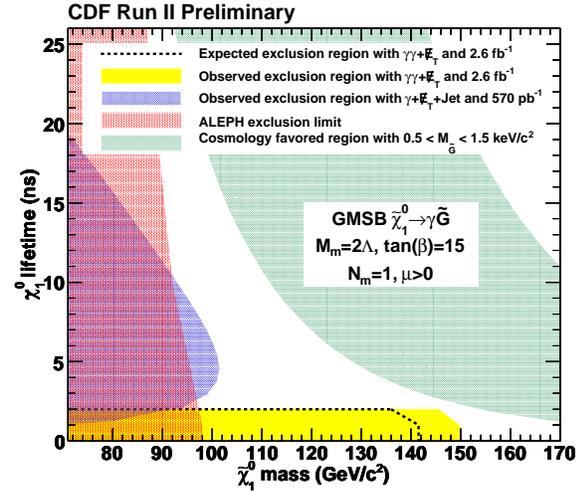}}
  \caption{The N-1 predicted kinematic distribution along with data after the
optimized
requirements
 are shown in Figure (b). There is no evidence for new physics and the data
is well modeled by backgrounds alone. In (b) the predicted and observed
exclusion region along with the limit from
    ALEPH/LEP~\cite{lep} and the $\gamma+\mett+jet$ delayed photon
analysis~\cite{delayedPRLD}.
    We have a mass reach of 141~GeV/$c^{2}$
    (predicted) and 149~GeV/$c^{2}$ (observed) at the lifetime up to 1~ns. The
green shaded band shows the parameter space where $0.5<m_{\Gravitino}<1.5~{\rm
keV}/c^{2}$, favored in cosmologically consistent
models~\cite{cosmology}.}
  \label{fig:exclusion}
\end{figure}

We have also re-optimize the sample for the GMSB model~\cite{gmsb}. In this
model, all SUSY pair production decays to two neutralinos, the
next-to-lightest SUSY particle, each of which then decays to a photon and a
gravitino, the lightest SUSY particle ($\none\to\gamma\grav$). We thus have two
photons, \mett, and other high-$E_{T}$ objects in the final state. Using
optimal set of cuts (\mett-significance, $\HT$, and $\DPHI$ between two photons)
we set the world's best limit on the \none mass of 149~GeV/$c^2$ at lifetime
below 1~ns. The results are shown in Fig.~\ref{fig:exclusion}.

\section{Search for Anomalous Production of Events with $\gamma$, jet, $b$-quark
jet, and \mett\ }
We search for new physics in the inclusive $\gamma b j\mett$ channel using 2.0
fb$^{-1}$ of data. We select an enhanced sample of events collected by an
inclusive isolated photon trigger. We require a central ($|\eta|<1.1$) photon
with $E_{T}>25$~GeV, two jets with $|\eta|<2.0$ and $E_{T}>15$~GeV, at least one
of which is identified as originating from a $b$-quark ($b$-tagged), and \mett\
greater than 25~GeV.

\begin{figure}[htbp]
  \centering
 {
\includegraphics[width=1.\linewidth]{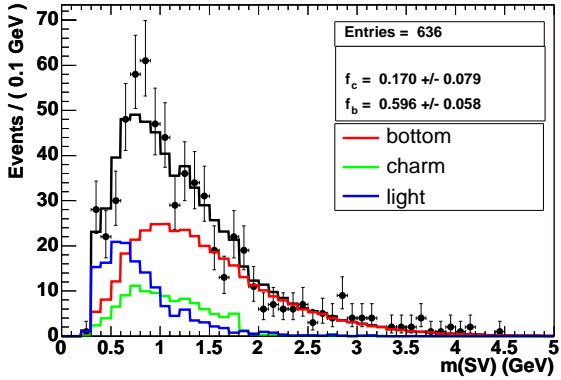}}
  \caption{The secondary vertex mass fit in events containing standard photons
for the search sample.}
  \label{fig:fit}
\end{figure}

The backgrounds are  misidentified photons (``misidentified $\gamma$''), true
photon plus light quark jet misidentified as heavy flavor (``true $\gamma$,
misidentified $b$''), true photon plus true $b$-tagged jet (``$\gamma b$''), and
true photon plus $c$-quark jet (``$\gamma c$''). The misidentified $\gamma$
background is estimated from the data sample by using cluster-shape variables
from the CES and hit rates in the CPR (the CES/CPR method). This technique
allows the determination of the number of photon candidates in the sample that
are actually misidentified jets as well as the corresponding shapes of the
distributions of kinematic variables.

The true $\gamma$,
misidentified $b$ background is estimated by first selecting events we then
apply the true-photon weight (the probability that a photon candidate is a
photon) determined using the CES/CPR method and the heavy-flavor mistag rate.
Because the CES/CPR method and the mistag parameterization provide
event-by-event weights, we are able to determine the shapes of kinematic
distributions as well as the number of events for this background.

We estimate the $\gamma b$ and $\gamma c$ backgrounds by generating MC events.
We obtain the overall normalizations of these backgrounds by fitting the
secondary vertex mass distribution of the tagged jets, $m(SV)$, to templates
built from the mass distributions of the expected SM components.
We first subtract the contribution due to misidentified photons by applying the
CES/CPR method to obtain the number of misidentified photon events. We then
estimate the fraction of heavy flavor in events with a misidentified photon by
fitting the secondary vertex mass distribution in a sample enriched with jets
faking photons. We then subtract the number of events containing a misidentified
photon and heavy flavor from the number of events obtained from the standard
photon sample fit to obtain the number of $\gamma b$ and $\gamma c$ events. 
Figure~\ref{fig:fit} shows the result of a secondary vertex fit performed on the
search region using the templates to extract the fraction of $b$-jet and $c$-jet
events.

\begin{figure}[htbp]
  \centering
\subfigure[]{
\includegraphics[width=1.\linewidth]{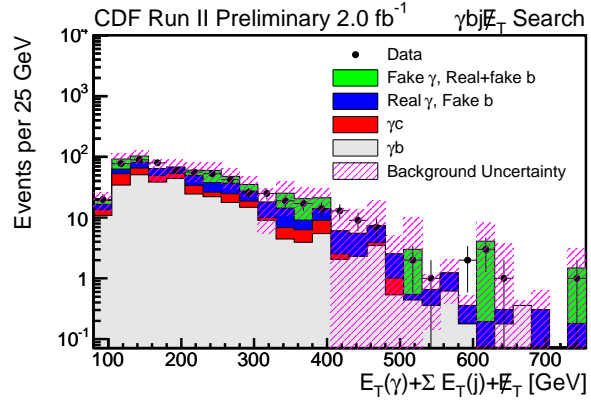}}
 \subfigure[]{
\includegraphics[width=1.\linewidth]{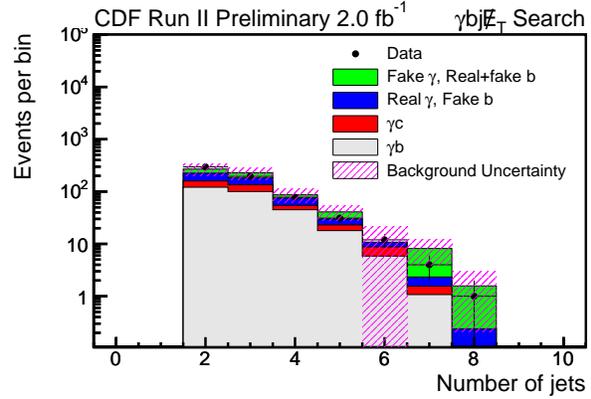}}
  \caption{In (a) the $\HT$ in the sample of a photon, a $b$-tagged jet, a second jet,
and \mett. In (b) the number of jets in the sample.}
  \label{fig:gbjmet}
\end{figure}

The total background prediction is 607$\pm$74$\pm$86, where the first
uncertainty is statistical and the second systematic. The observed number of
events is 617, consistent with the SM background predictions.
Figure~\ref{fig:gbjmet} shows the $\HT$ and the number of jets distributions in the sample of a photon,
a $b$-tagged jet, a second jet and $\mett$. This result is now
published~\cite{gbjmet}.

\section{Search for a Fermiophobic Higgs Boson in $\gamma\gamma$}
The SM prediction for the Higgs, $h\to\gamma\gamma$ branching ratio is extremely
small. However, in ``fermiophobic'' models, where the coupling of the Higgs
boson to fermions is highly suppressed, the diphoton decay can be greatly
enhanced. Since for this fermiophobic case, the diphoton final state dominates
at low Higgs boson masses the diphoton final state becomes the preferred search
channel. 

We select a diphoton sample from 3.0 fb$^{-1}$ of data, filtered by diphoton
trigger. We then require both photons to be located within central region
($|\eta|<1.05$), referred to as ``central-central region'', 
or one photon to be in this region and the other photon in
plug region ($1.2<|\eta|<2.8$), referred to as ``central-forward region''.
Individual photons are required to have
$E_{T}>15$~GeV, while the diphoton pair is required to have mass of
$m_{\gamma\gamma}>30$~GeV/$c^2$. However, the fermiophobic HIggs boson is only
produced at a non-negligible rate in association with a $W$/$Z$ boson or via
vector boson fusion process. Since associated production dominates the
production process the optimization was performed on the basis of the associated
production process alone. A selection based on the following observables was
optimized: diphoton transverse momentum ($p^{\gamma\gamma}_{T}$), transverse
momentum of the second highest $p_T$ jet ($p^{j2}_{T}$) for hadronic decays of
$W/Z$, and missing transverse energy (\mett) or transverse momentum of an
isolated track ($p^{iso}_{T}$) for leptonic decays of $W/Z$. 
A variety of sets of these requirements which would select
evidence of the $W$ or $Z$ boson were carried out, but the only single
requirement that the diphoton transverse momentum ($p^{\gamma\gamma}_{T}$) be
greater than 75~GeV/$c$ is approximately as sensitive as any combination of the
other selection requirements. With this requirement on $p^{\gamma\gamma}_{T}$,
roughly 30\% of the signal remains while more than 99.5\% of the background is
removed.

\begin{figure}[htbp]
  \centering
  \subfigure[]{
    \includegraphics[width=.48\linewidth]{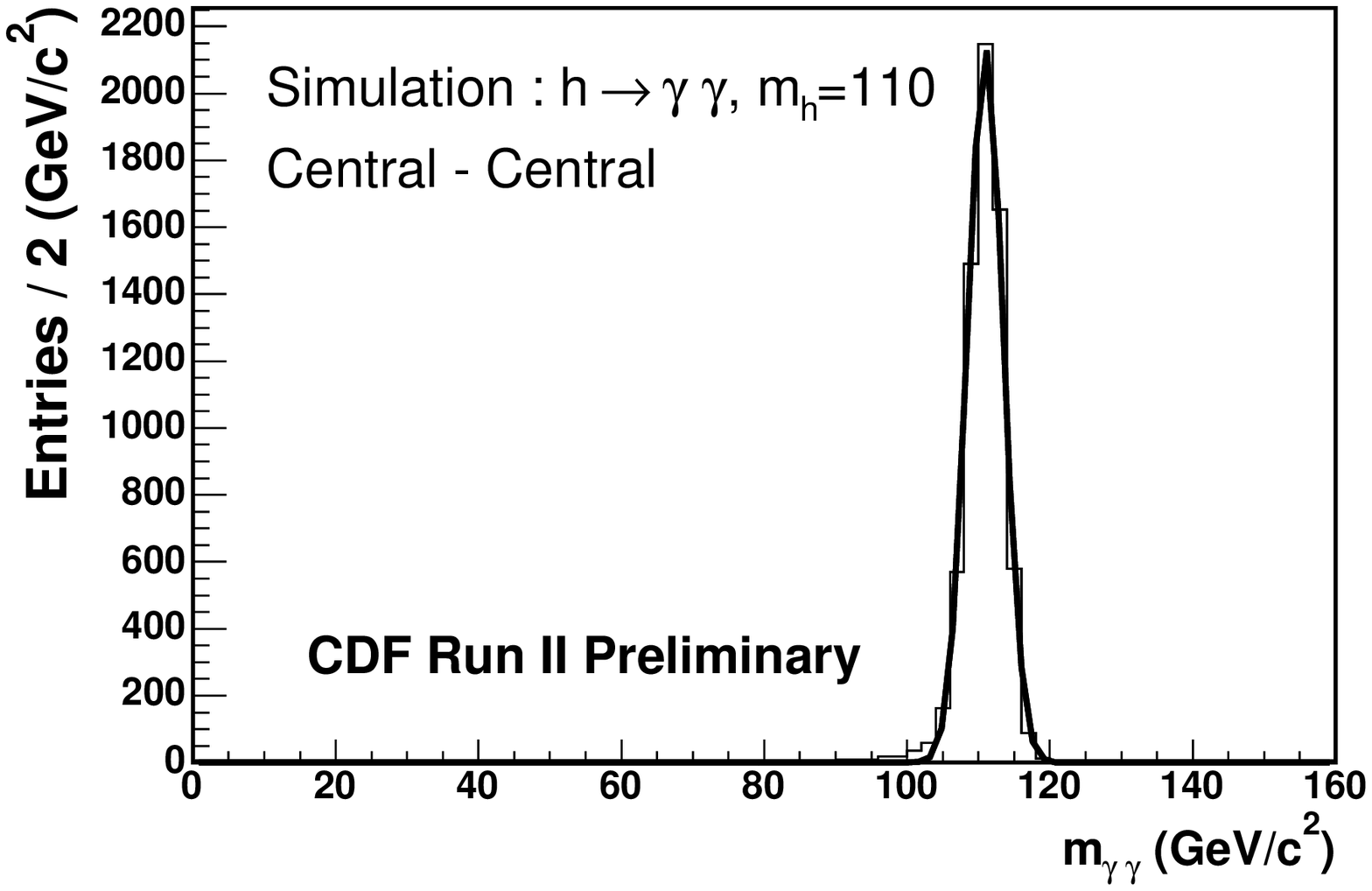}}
  \subfigure[]{
    \includegraphics[width=.48\linewidth]{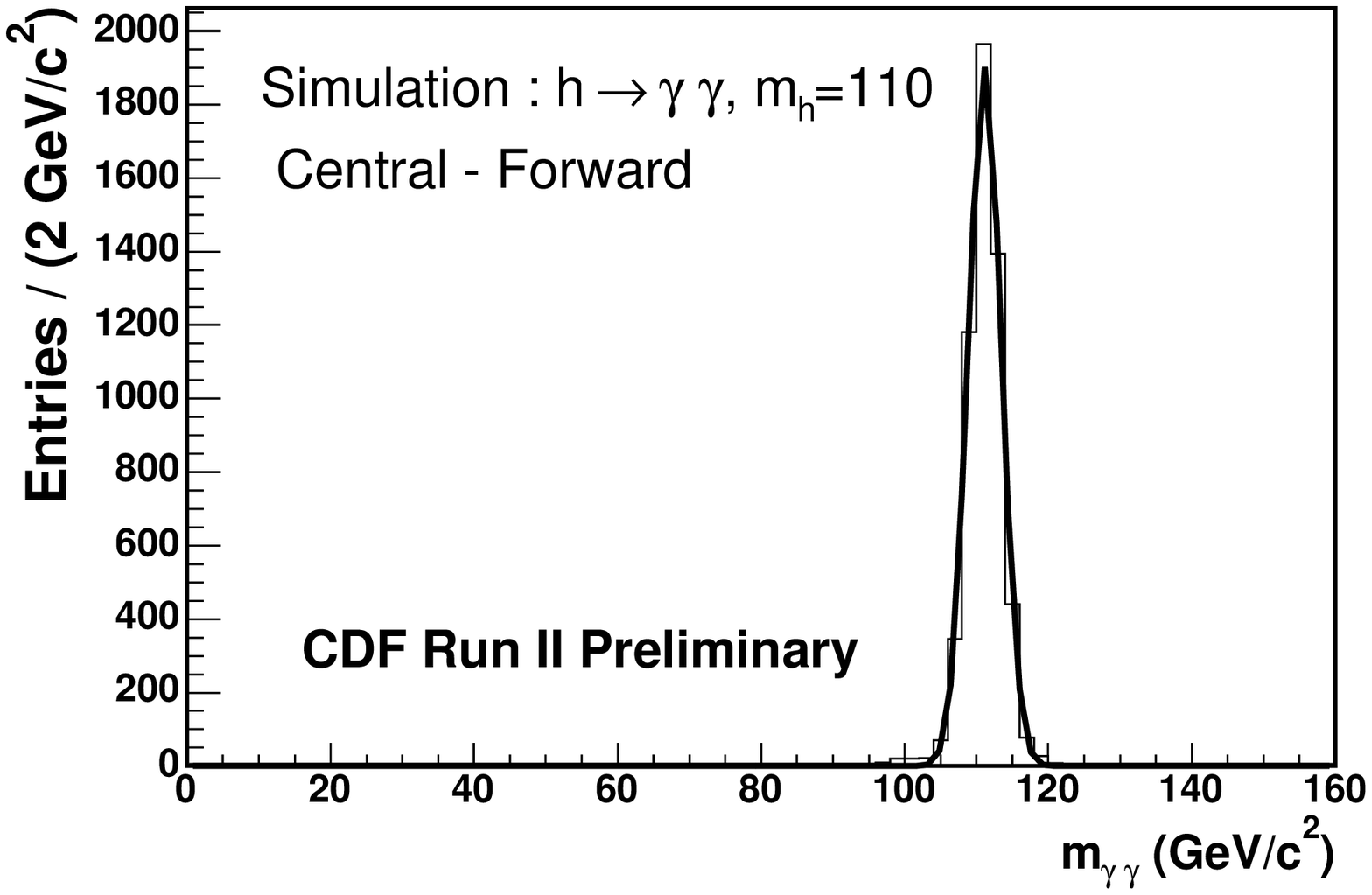}}
  \subfigure[]{
    \includegraphics[width=1.\linewidth]{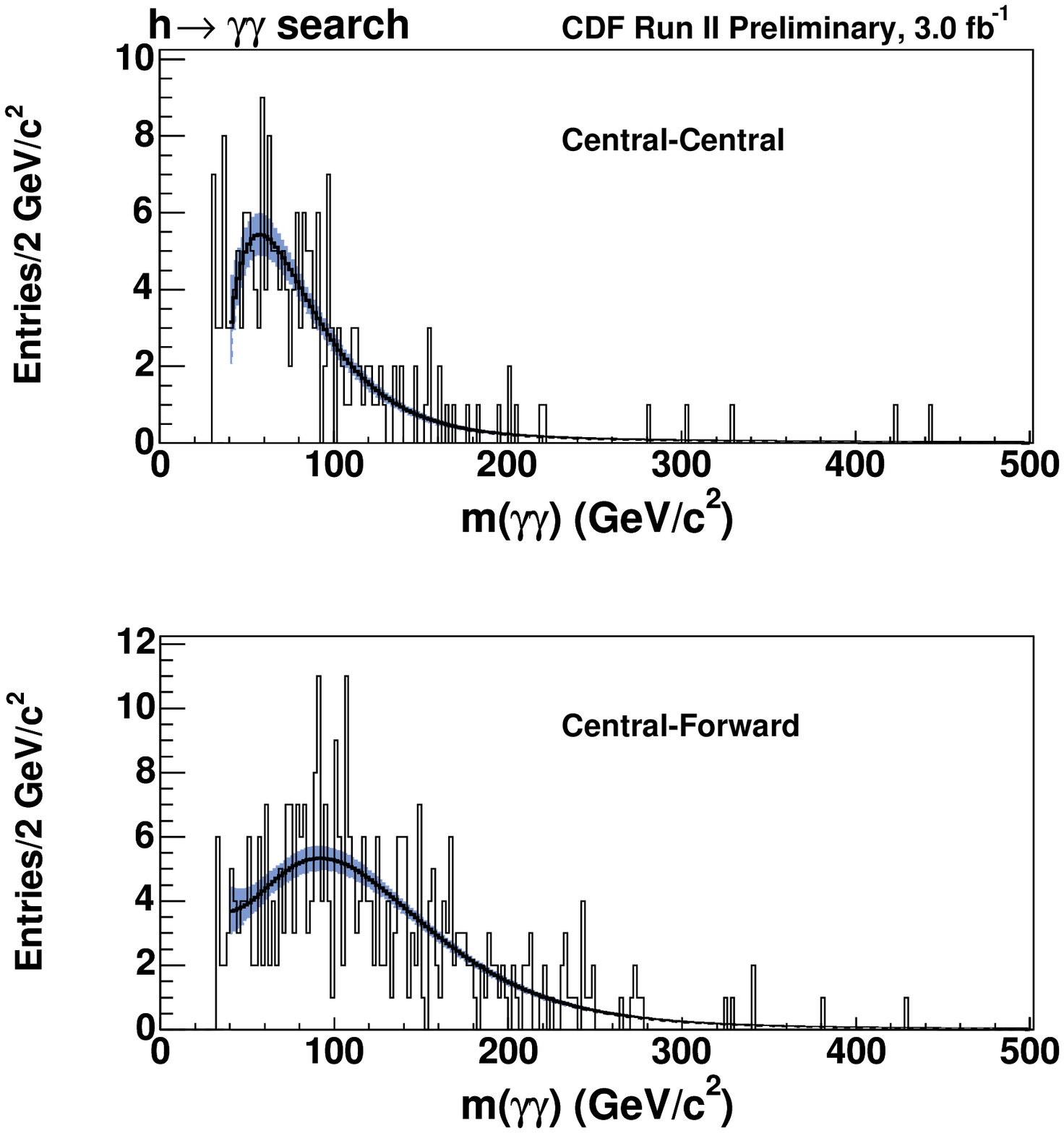}}
  \caption{In (a) and (b) the expected shapes of invariant Higgs mass of the
signal from simulation are
shown for central-central and central-forward regions, respectively.
In (c) the invariant mass distribution of central-central (top) and
central-forward (bottom) photon pairs after the requirement of
$p_{T}^{\gamma\gamma}>75$~GeV/$c$ with the fit to the data for the hypothesis
of a $m_{h}=100$~GeV/$c^2$.}
  \label{fig:dipho_mass}
\end{figure}

\begin{figure}[htbp]
  \centering
  \subfigure[]{
  \includegraphics[width=1.\linewidth]{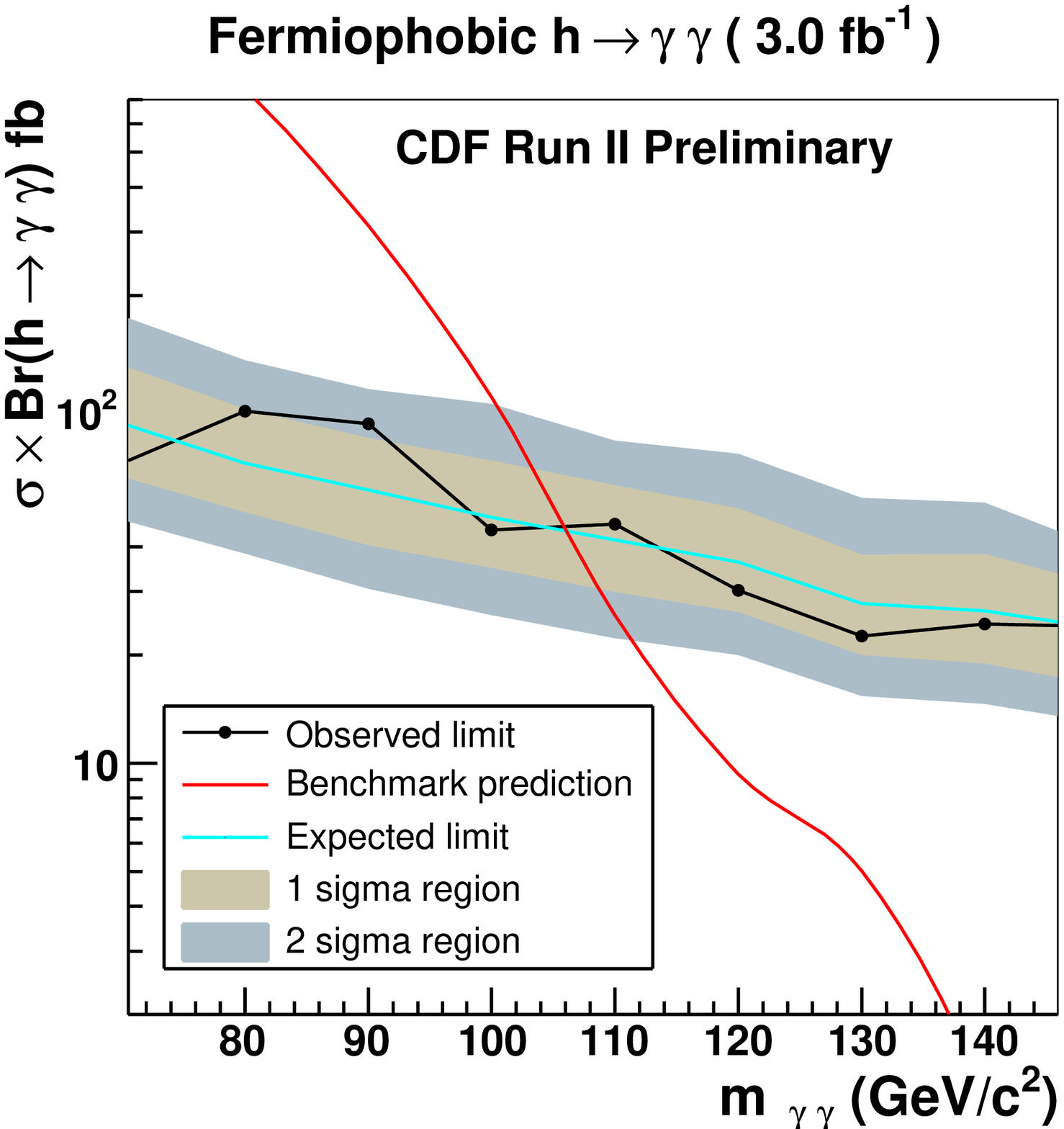}}
 \subfigure[]{
  \includegraphics[width=1.\linewidth]{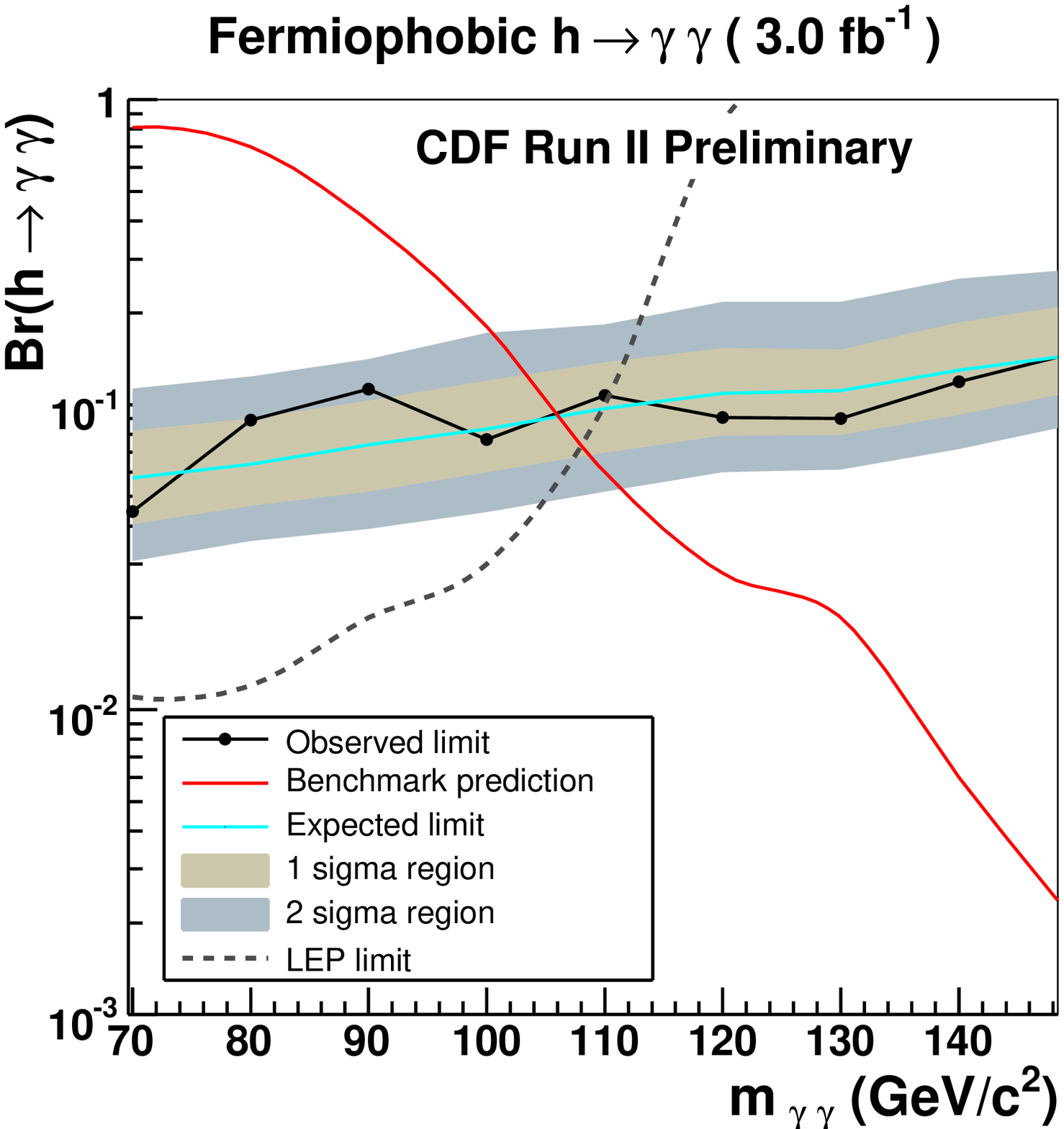}}
  \caption{The 95\% C.L. upper limit on the production cross section (a) and the
branching fraction (b) for the fermiophobic Higgs boson decay to diphotons, as a
function of $m_{h}$.}
  \label{fig:higgs}
\end{figure}

The decay of a Higgs boson into a diphoton pair appears as a very narrow peak in the
invariant mass distribution of this diphoton pair. The diphoton mass resolution
determined from simulation is better than 3\% for the Higgs boson mass region,
as shown in Fig.~\ref{fig:dipho_mass}-(a) and (b).
To establish the level of signal and background, we fit the diphoton mass 
spectrum to a background shape and the signal shape from a MC sample.
The resulting invariant mass distributions of central-central and
central-forward diphoton pair is
shown in Fig.~\ref{fig:dipho_mass}-(c). 

No evidence of such a resonance appears
in the data so we set the 95\% C.L. upper limits both on the production cross
section ($\sigma\times B(h\to\gamma\gamma)$) and the branching fraction for the fermiophobic
Higgs boson decay to diphotons as a function of this Higgs mass ($m_{h}$) and
exclude this type of Higgs boson mass up to 106~GeV/$c^2$, as shown in
Fig.~\ref{fig:higgs}. This result is now published~\cite{fhiggs}

\begin{acknowledgments}
I would like to thank the authors: R.~Culbertson, H.~Frisch, D.~Krop, C.~Pilcher, S~Wilbur,
S.~Yu, A.~Pronko, M.~Goncharov, D.~Toback for their analyses and help on the
talk. 
\end{acknowledgments}
\bigskip 

\end{document}